\documentclass[aps,prd,reprint,10pt,nofootinbib]{revtex4-1}
\usepackage{amsmath}
\usepackage{amsfonts}
\usepackage{graphicx}
\usepackage{subcaption}
\usepackage{caption}
\usepackage{comment}
\usepackage{toolbox}
\usepackage{xcolor}
\usepackage{braket}
\usepackage{hyperref}

\begin{document}

\title{\LARGE $W$-Boson Mass Anomaly from Scale Invariant 2HDM}

\author{Karim Ghorbani}
\affiliation{Physics Department, Faculty of Sciences, Arak University, Arak 38156-8-8349, Iran}
\author{Parsa Ghorbani}
\affiliation{Physics Department, Faculty of Science, Ferdowsi University of Mashhad, Iran}

\begin{abstract}
The recently reported measurement of the $W$-boson mass by CDF-II collaboration is significantly heavier than that of the Standard Model prediction. We study this anomaly in the scale invariant Two-Higgs-Doublet-Model (SI-2HDM) with a $\mathbb{Z}_2$  symmetry to avoid the flavor-changing-neutral-current (FCNC). In this scenario the Higgs particle is the classically massless scalon in the SI-2HDM gaining its mass by radiative corrections, hence being naturally light. Moreover, because of the scale symmetry the model is more predictive respect to the generic 2HDM. 
We show that the oblique parameters depending on the masses of the charged and CP-even (CP-odd) neutral scalar components of the SI-2HDM denoted respectively by $M_{H^\pm}$ and  $M_h$ ($M_A$), are large enough to accommodate the $W$-boson mass anomaly in the model. There are as well viable regions in the mass spectrum of the SI-2HDM that evade the experimental bounds from colliders on the charged Higgs and neutral scalars. 
\end{abstract}

\maketitle


\section{Introduction}

The Standard Model (SM) prediction for the $W$ gauge boson mass is $M_W^\text{SM}=80.357\pm 0.006$ GeV \cite{ParticleDataGroup:2020ssz}. This theoretical result was confirmed by the LEP \cite{ALEPH:2013dgf}, the ATLAS \cite{ATLAS:2017rzl}, the LHCb \cite{LHCb:2021bjt}, and by the CDF and D0 collaborations on Tevatron \cite{CDF:2013dpa}, giving rise to the PDG average $M_W^\text{PDG}=80.379\pm 0.012$ GeV. However, recently the CDF II collaboration \cite{CDF:2022hxs} reported a remarkable result for the measurement of the $W$ gauge boson mass; $M_W^\text{CDF}=80.43335\pm 0.0094$ GeV which is $7\sigma$ standard deviation away from the SM prediction. 
In order to explain this anomaly the addition of extra fields seems inevitable, therefore it could be a window towards New Physics (NP) beyond the SM. Different models have been considered so far to interpret the anomaly  \cite{Fan:2022dck,Lu:2022bgw,Athron:2022qpo,Yuan:2022cpw,Strumia:2022qkt,Yang:2022gvz,deBlas:2022hdk,Du:2022pbp,Tang:2022pxh,Cacciapaglia:2022xih,Blennow:2022yfm,Sakurai:2022hwh,Fan:2022yly,Liu:2022jdq,Lee:2022nqz,Cheng:2022jyi,Bagnaschi:2022whn,Paul:2022dds,Bahl:2022xzi,Asadi:2022xiy,DiLuzio:2022xns,Athron:2022isz,Gu:2022htv,Babu:2022pdn,Heo:2022dey,Du:2022brr,Cheung:2022zsb,Crivellin:2022fdf,Endo:2022kiw,Biekotter:2022abc,Balkin:2022glu,Han:2022juu,Ahn:2022xeq,Kawamura:2022uft,Zheng:2022irz,Ghoshal:2022vzo,Perez:2022uil,Mondal:2022xdy,Zhang:2022nnh,Borah:2022obi,Chowdhury:2022moc,Arcadi:2022dmt,Cirigliano:2022qdm,Heckman:2022the,Kanemura:2022ahw,Popov:2022ldh}.
In this work we confront the scale invariant two Higgs doublet model (SI-2HDM) with the reported $W$-boson mass anomaly by the CDF II collaboration. Our motivation to study this model is as follows. Within the mass spectrum of the SI-2HDM, there is a classically massless scalon field which we take as the SM Higgs particle. As the scalon gains mass  through only radiative corrections, therefore the SM Higgs becomes naturally as light as the EW scale. Moreover, the free parameters in the SI-2HDM due to scale symmetry are only two which is clearly less than the corresponding generic 2HDM model, making the scenario much more predictive. In the mass spectrum, apart from the SM Higgs, there are two charged scalars, a CP-even and a CP-odd scalar denoted by $H^\pm$, $h$ and $A$ respectively. All these scalars contribute in the oblique parameters. We show there are viable scalar masses that give rise to the desired values for the oblique parameters appropriate for the reported $W$-boson mass anomaly. We will constrain more the parameter space from colliders on the mass of the charged Higgs. 

The content of this paper is arranged as follows. We first introduce the SI-2HDM in the next section. Then in section \ref{const}, we will discuss the theoretical and experimental constraints on the model. In section \ref{res} we present the results and will determine for what scalars masses the $W$-boson mass anomaly can be explained. We conclude in section \ref{conc}.     

\section{Scale Invariant 2HDM}\label{model}
We choose a SI-2HDM which is a minimal and highly predictive model due to the scale symmetry restrictions. 
The scale symmetry means that all  dimensionful parameters e.g. the mass parameters are absent in the theory. In the SM if we ignore the Higgs mass term $\mu^2_{1} |\Phi_1|^2$, with $\Phi_1$ being the Higgs doublet, the SM action becomes invariant under scale transformations. Alternatively, in order to make the theory scale invariant the dimensionful mass parameter $\mu^2_{1}$ can be replaced by another doublet $\Phi_2$, to form a bi-quadratic term in the potential introducing a new dimensionless coupling  $\lambda_{3}$: $\lambda_{3} |\Phi_1|^2  |\Phi_2|^2$. A general potential of 2HDM 
consisting of two  doublets $\Phi_1$ and $\Phi_2$ is as the following (for more detail on SI-2HDM see \cite{Lee:2012jn})
\begin{equation}\label{vtree}
\begin{split}
V=&\frac{\lambda_1}{2} (\Phi^\dagger_1 \Phi_1)^2
+\frac{\lambda_2}{2} (\Phi^\dagger_2 \Phi_2)^2
+\lambda_3 (\Phi^\dagger_1 \Phi_1)(\Phi^\dagger_2 \Phi_2)\\
&+\lambda_4 (\Phi^\dagger_1 \Phi_2)(\Phi^\dagger_2 \Phi_1)+\frac{\lambda_5}{2} \left[ (\Phi^\dagger_1 \Phi_2)^2+ \text{h.c.} \right]
\end{split}
\end{equation}
where all dimensionless couplings in Eq. (\ref{vtree}) are real, and
\begin{equation}\label{phii}
\Phi_i=\begin{pmatrix}
\varphi_i^+  \\
(v_i+h_i+iz_i)/\sqrt{2}
\end{pmatrix}.
\end{equation}
Depending on how the doublets interact with the SM quarks and leptons, the type of the 2HDM is determined. In the 2HDM type I (2HDM-I), the SM fermions interact with only one doublet, here with $\Phi_1$
\begin{equation}\label{lyuk1}
\mathcal{L}_Y \supset y^u_{ij}  \bar Q_{L_i} \tilde{\Phi}_1 u_{R_j} 
+  y^d_{ij}  \bar Q_{L_i} \Phi_1 d_{R_j} 
+ y^l_{ij}  \bar E_{L_i} \Phi_1 e_{R_j}+\text{h.c.}
\end{equation} 
while in the 2HDM type II (2HDM-II), both doublets interact with the SM fermions as
\begin{equation}\label{lyuk2}
\mathcal{L}_Y \supset y^u_{ij}  \bar Q_{L_i} \tilde{\Phi}_1 u_{R_j} 
+  y^d_{ij}  \bar Q_{L_i} \Phi_2 d_{R_j} 
+ y^l_{ij}  \bar E_{L_i} \Phi_2 e_{R_j}+\text{h.c.}
\end{equation} 
where $\tilde{\Phi}_i=i\sigma_2 \Phi_1^*$. 
In our notation, the doublet $\Phi_1$ includes the SM Higgs component. To avoid the flavor changing neutral currents (FCNCs), we have imposed a $\mathbb{Z}_2$ symmetry in Eqs. (\ref{vtree}), (\ref{lyuk1}) and  (\ref{lyuk2}) under which $\Phi_1\to -\Phi_1$, $\Phi_2\to \Phi_2$ and $u_R\to -u_R$. That is the reason for we have not included the terms $(\Phi_1^\dagger \Phi_1)(\Phi_1^\dagger \Phi_2)$ and $(\Phi_2^\dagger \Phi_2)(\Phi_1^\dagger \Phi_2)$ in Eq. (\ref{vtree}) as they violate the $\mathbb{Z}_2$ symmetry.

The scale symmetry is broken through radiative corrections {\it \`{a} la} Coleman and Weinberg (CW) \cite{Coleman:1973jx}. Gildener and Weinberg (GW) \cite{Gildener:1976ih} pointed out that in a scale invariant theory with multiple scalars there is always one classically massless state dubbed as {\it scalon}. The scale symmetry is broken via radiative corrections and the classically massless scalon gets massive. Following GW, the radiative corrections are calculated along the {\it flat direction} where the potential at minimum is vanishing. 
The SI-2HDM potential in Eq. (\ref{vtree}) can be written in the form proposed by GW as $V=1/8\lambda_1\phi_1^4+1/8\lambda_2\phi_2^4+1/8\lambda_{345} \phi_1^2 \phi_2^2$ where $\Phi_1^\dagger\equiv (0~\phi_1)/\sqrt{2}$ and $\Phi_2^\dagger\equiv (0~\phi_2)/\sqrt{2}$ and $\lambda_{345}=\lambda_3+\lambda_4+\lambda_5$.
We now define the flat direction as the unit vector $(n_1,n_2)$ for which $\phi_1=\phi n_1$ and $\phi_2=\phi n_2$,
where $\phi^2=\phi_1^2+\phi_2^2$ is the radial field. Then the potential and the minimum along this flat direction is vanishing if  
\begin{equation}
\sqrt{\lambda_1 \lambda_2}+\lambda_{345}=0, 
\end{equation} 
and 
\begin{equation}
n_1= \frac{\sqrt{\lambda_2}}{\sqrt{\lambda_1}+\sqrt{\lambda_2}} ~~~~
n_2= \frac{\sqrt{\lambda_1}}{\sqrt{\lambda_1}+\sqrt{\lambda_2}}\,.
\end{equation}
Both doublets develop non-zero vacuum expectation values (VEV)
\begin{equation}
 v_1\equiv \braket{\phi} n_1=v n_1 \hspace{1cm} 
 v_2\equiv \braket{\phi} n_2=v n_2
\end{equation}
where $v^2=v_1^2+v_2^2$ and the ratio is given by the angle
$\tan\beta = v_1/v_2 \equiv n_1/n_2$. Now a rotation in the space of doublets $(\Phi_1,\Phi_2)$ by the angle $\beta$ brings us to a new basis $(H_1,H_2)$ where  $\braket{H_1}=v$ and $\braket{H_2}=0$. The new doublets $H_1$ and $H_2$ can be written as 
\begin{equation}\label{H1H2}
H_1=\begin{pmatrix}
G^+  \\
(v+H+iG^0)/\sqrt{2}
\end{pmatrix} ~~~H_2=\begin{pmatrix}
H^+  \\
(h+iA)/\sqrt{2}
\end{pmatrix} 
\end{equation}
where we assume that $H_1$ is the SM doublet with $v=246$ GeV being the Higgs VEV, and $H_2$ is the new doublet. Three components $G^\pm$ and $G^0$ in $H_1$ are gauged away. Substituting Eq. (\ref{H1H2}) (after rotating by angle $\beta$ to the $(\Phi_1,\Phi_2)$ basis) in Eq. (\ref{vtree})
the mass terms in the tree-level potential are given as 
\begin{equation}
\begin{split}
 V_\text{mass}= (H^-~G^-) \begin{pmatrix}
                          m^2_{H^\pm}(\phi) & 0\\
                          0  & 0
                         \end{pmatrix} \begin{pmatrix}
                         H^+ \\
                         G^+
                         \end{pmatrix} \\
                         +\frac{1}{2}(A~G^0)\begin{pmatrix}
                         m^2_A(\phi) & 0\\
                         0 & 0
                         \end{pmatrix}
                         \begin{pmatrix}
                         A \\
                         G^0
                         \end{pmatrix}\\
                         +\frac{1}{2}(h~H) \begin{pmatrix}
                         m^2_h(\phi) & 0 \\
                         0 & 0
                         \end{pmatrix} 
                         \begin{pmatrix}
                          h \\
                          H
                         \end{pmatrix},
\end{split}
\end{equation}
leading to the field-dependent mass eigenvalues
\begin{equation}\label{diagmass}
\begin{split}
&m_H^2=0 \hspace{1.25in} \, m_h^2(\phi)=\sqrt{\lambda_1 \lambda_2}\, \phi^2\\
&m_{H^\pm}^2(\phi)=\frac{1}{2}(\sqrt{\lambda_1 \lambda_2}+\lambda_3)\phi^2 ~~~~ m_A^2(\phi)=-\lambda_5 \phi^2.
\end{split}
\end{equation}
We take the scalon $H$ as the Higgs field in the SM with $m_H\sim 125$ GeV. 

The classically scale symmetry is broken by radiative corrections. The one-loop effective potential is given by \cite{Coleman:1973jx}

\begin{equation}
 V_\text{eff}^\text{1-loop}=\sum_i n_i \frac{m_i^4(\phi)}{64\pi^2}\left( \ln\frac{m_i^2(\phi)}{\bar\mu^2} -C_i\right),
\end{equation}
where $i$ runs for all scalar, gauge boson and fermion fields in the model, $n_i$ is the number of degrees of freedom for each field e.g. $n_{H^\pm}=2$ and $n_t=-12$, $m_i(\phi)$ is the field-dependent mass, and $\bar\mu$ is the renormalization scale. The coefficients $C_i$ take the value $3/2$ ($5/6$) for scalars and fermions (gauge bosons). Using Eq. (\ref{diagmass}) for $m_i(\phi)$ the one-loop effective potential reads
\begin{equation}\label{effpot}
 V_\text{eff}^\text{1-loop}=A\phi^4+B\phi^4 \ln\frac{\phi^2}{\bar\mu^2},
\end{equation}
with 
\begin{subequations}
\begin{align}
&A=\sum_i n_i \frac{M_i^4}{64\pi^2}\left( \ln\frac{M_i^2}{v^2} -C_i\right),\\
&B=\sum_in_i\frac{M_i^4}{64\pi^2 v^4},
\end{align}
\end{subequations}
where $M_i\equiv m_i(v)$. Now the flatness condition for the one-loop effective potential in Eq. (\ref{effpot}), relates the renormalization scale $\bar\mu$ and $v$ as $\ln(\bar\mu/v)=A/2B+1/4$, hence the potential in the vacuum is $V_\text{eff}^\text{1-loop}(v)=-Bv^4/2$. 
The radiative correction to the scalon mass along the flat direction is given by the second derivative of the effective potential at the minimum $\phi=v$
\begin{equation}\label{mH2}
\begin{split}
 \delta m_H^2=&\frac{\partial^2V_\text{eff}^\text{1-loop}(\phi)}{\partial\phi^2}\Big\vert_{\phi=v}=8Bv^2\\
 &=\frac{1}{8\pi^2 v^2}(6M_W^4+3M_Z^4-12M_t^4\\
&+2M_{H^\pm}^4+M_h^4+M_A^4 ), 
 \end{split}
\end{equation}
where we have kept only the top quark contribution and discarded the negligible terms including light quarks and leptons. Note that the radiative corrections to the scalon mass in Eq. (\ref{mH2}) is positive at the electroweak scale provided that the new charged and neutral scalars are heavy enough to compensate the top quark negative contribution. By solving the renormalization group equations (RGE) in section \ref{perunit}, we will find the scale up to which the model stays valid. 

\section{Oblique Parameters}
The CDF-II measurement on the W-boson mass was performed using Fermilab Tevatron collider $8.8 \text{fb}^{-1}$ data. The oblique parameters $S$, $T$ and $U$ define the effects of radiative corrections from new physics \cite{Peskin:1990zt,Peskin:1991sw}. Fixing $U=0$, the global fit of the electroweak data with the CDF-II measurement gives rise to the following large central values \cite{Strumia:2022qkt}
\begin{equation}
 T_0=\left(0.1548, 0.0645\right),\hspace{1cm}  S_0=\left(0.0595 , 0.0833\right),
\end{equation}
with strong correlation $\rho_\text{ST}=0.95$.
In the SI-2HDM, the scalars $H^\pm, h$ and $A$ contribute in the oblique parameters which are given by
\begin{equation}\label{op}
\begin{split}
&S=\frac{1}{12\pi}\left[g(M_A^2,M_h^2)+\ln\frac{M_A M_h}{M_{H^\pm}^2} \right],\\
&T=\frac{1}{16\pi}\frac{1}{s_W^2 c_W^2 M_Z^2}[f(M_A^2,M_{H^\pm}^2)\\
&~~~~~~~~~~~~~~~~~~~~~~~~~~~~~~+ f(M_h^2,M_{H^\pm}^2)-f(M_A^2,M_h^2)],\\
&U=\frac{1}{12\pi}\left[g(M_A^2,M_{H^\pm}^2)+g(M_h^2,M_{H^\pm}^2)-g(M_A^2,M_h^2) \right],
\end{split}
\end{equation}
where the functions $f(x,y)$ and $g(x,y)$ are defined as
\begin{equation}
\begin{split}
&f(x,y)=\frac{x+y}{2}-\frac{xy}{x-y}\ln\frac{x}{y},\\
&g(x,y)=-\frac{5}{6}+\frac{2xy}{(x-y)^2}+\frac{(x+y)(x^2-4xy+y^2)}{2(x-y)^3}\ln\frac{x}{y},
\end{split}
\end{equation}
with $x,y>0$.
The electroweak oblique parameters $S$ and $T$ are constrained as \cite{Lee:2012jn}
\begin{equation}\label{STconst}
\begin{split}
&\frac{(S- S_0)^2}{\sigma_\text{S}^2}+\frac{(T- T_0)^2}{\sigma_\text{T}^2}\\
&~~~~~~~~~-2\rho_\text{ST}\frac{(S- S_0)(T- T_0)}{\sigma_\text{S}\sigma_\text{S}} \leq R^2 (1-\rho^2_\text{ST})
\end{split}
\end{equation}
where $ S_0$ and $T_0$ are the central values and $\sigma_\text{S}$ and $\sigma_\text{T}$ are the standard deviations for the oblique parameters $S$ and $T$, respectively. The parameter $R^2$ determines the confidence level (CL).

\section{Perturbativity and unitarity}\label{perunit}
\begin{figure}
 \includegraphics[scale=.9]{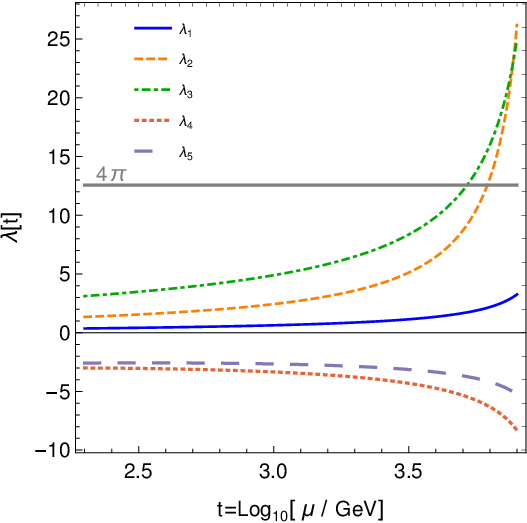}
 \caption{The running of the quartic couplings is shown. The couplings $\lambda_2$ and $\lambda_3$ hit a Landua pole around the scale $4$ TeV.}
 \label{rge}
\end{figure}

\begin{figure}
 \includegraphics[scale=.9]{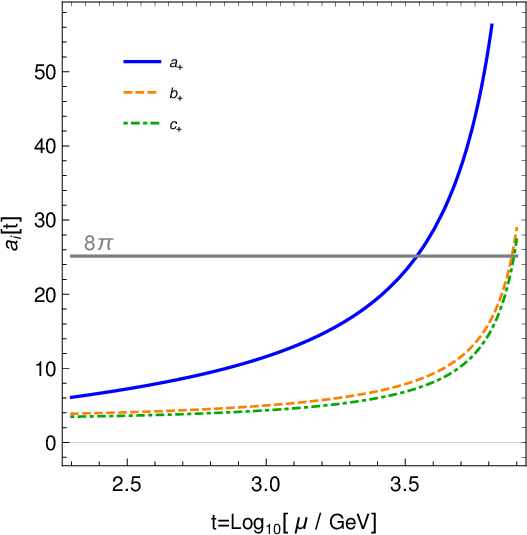}
 \caption{The running of the eigenvalues of the two-particle scattering matrix is shown. The unitarity is violated around the scale $4$ TeV.}
 \label{unitarity}
\end{figure}

The perturbativity conditions on the quartic couplings of the potential at a given scale $\Lambda$ require that $|\lambda_i(\Lambda)| < 4\pi$.
This imposes upper limits on the couplings at an arbitrary scale.
Applying the renormalization group equations we find beta functions at one loop for the quartic couplings \footnote{We have used the package SARAH, \cite{Staub:2013tta} to obtain the beta functions.}
\begin{equation}\label{betafun}
\begin{split}
 (16\pi^2) \beta_{\lambda_1}  &= 9 g_2^2 \lambda_1 + 24 \lambda_1^2 + \lambda_5^2+ 2 \lambda_3^2 + 2 \lambda_3 \lambda_4 + \lambda_4^2 \\
 &+\frac{27}{200} g_1^4 + \frac{9}{8} g_2^4   
 + \frac{9}{20} g_1^2 g_2^2 - \frac{9}{5} g_1^2 \lambda_1   \,, \\
 (16\pi^2) \beta_{\lambda_2}  &=
24 \lambda_2^2 + 2 \lambda_3^2 + 2 \lambda_3 \lambda_4 + \lambda_4^2 + \lambda_5^2 
 + 12 \lambda_2 y_b^2 \\
 &+ 12 \lambda_2 y_t^2 + 4 \lambda_2 y_\tau^2
 + \frac{27}{200} g_1^4 + \frac{9}{20} g_1^2g_2^2 + \frac{9}{8} g_2^4 \\
 &+ \frac{9}{5} g_1^2\lambda_2 - 9 g_2^2 \lambda_2 - 6 y_b^4  
 - 6 y_t^4 - 6 y_\tau^4 \,, \\
 (16\pi^2) \beta_{\lambda_3}  &=
  4 \lambda_3^2 + 4 \lambda_1 \lambda_4 + 4  \lambda_2 \lambda_4 
  +2 \lambda_4^2 + 2 \lambda_5^2 + 12 \lambda_1 \lambda_3 \\
  &+ 12 \lambda_2 \lambda_3 +  + 6 \lambda_3 y_b^2 
   +2 \lambda_3 y_\tau + 6 \lambda_3 y_t^2 \\
  &+ \frac{27}{100} g_1^4  - \frac{9}{10} g_1^2 g_2^2 + \frac{9}{4} g_2^4 - 
  \frac{9}{5} g_1^2 \lambda_3 - 9 g_2^2 \lambda_3 \,, \\
(16\pi^2) \beta_{\lambda_4}  &=
    4 \lambda_4^2 + 8 \lambda_5^2 + 4 \lambda_1 \lambda_4 
    +4 \lambda_2 \lambda_4 + 8 \lambda_3 \lambda_4
    +6 \lambda_4 y_b^2   \\
    & + 2 \lambda_4 y_\tau^2 
      - \frac{9}{5} g1^2 \lambda_4 - 9 g_2^2 \lambda_4 
      + 6 \lambda_4 y_t + \frac{9}{5} g_1^2 g_2^2 \,, \\
(16\pi^2) \beta_{\lambda_5}  &=
   4 \lambda_1 \lambda_5    + 4 \lambda_2 \lambda_5 + 8 \lambda_3 \lambda_5 
   + 12 \lambda_4 \lambda_5  + 6 \lambda_5 y_b^2 \\
   &+ 2 \lambda_5 y_\tau + 6 \lambda_5 y_t^2 
   -\frac{9}{5} g_1^2 \lambda_5 - 9 g_2^2 \lambda_5 \,.
\end{split}
\end{equation}
The beta functions for the Yukawa couplings are given by 
\begin{equation}
 \begin{split}
&(16\pi^2)\beta_{y_t} = ( \frac{9}{2} y_t^2 + \frac{3}{2} y_b^2  + y_\tau^2
   -\frac{17}{20} g_1^2 -\frac{9}{4} g_2^2  -8 g_3^2 )y_t \,, \\
&(16\pi^2)\beta_{y_b} = ( \frac{9}{2} y_b^2 + \frac{3}{2} y_t^2  + y_\tau^2
   +\frac{1}{4} g_1^2 -\frac{9}{4} g_2^2  -8 g_3^2 )y_b \,, \\
&(16\pi^2)\beta_{y_\tau} = ( \frac{9}{2} y_\tau^2 + 3 y_t^2  + 3 y_b^2
   +\frac{9}{4} g_1^2 -\frac{9}{4} g_2^2 )y_\tau \,.
   \end{split}
\end{equation}
There are also strong constraints from unitarity conditions on the tree-level scattering amplitude with the addition of the new scalars. To extract these bounds one has to obtain the eigenvalues of a two-particle scattering matrix by considering all the $2 \to 2$ scattering processes with various initial and final states \cite{Gorczyca:2011he,Lee:1977eg,Ginzburg:2005dt,Bhattacharyya:2013rya,Chakrabarty:2014aya}.
These eigenvalues,  say $\mathcal{M}_i$, should satisfy the upper limit $|\mathcal{M}_i| \leq 8\pi$. The analytical expressions for these eigenvalues are obtained as
\begin{equation}
\begin{split}
 a_{\pm} &= \frac{3}{2}(\lambda_1 + \lambda_2)\pm \sqrt{\frac{9}{4}(\lambda_1-\lambda_2)^2+(2\lambda_3+\lambda_4)^2} \,, \\
 b_{\pm} &= \frac{1}{2}(\lambda_1 + \lambda_2)\pm \sqrt{\frac{1}{4}(\lambda_1-\lambda_2)^2+\lambda_4^2} \,, \\
c_{\pm} &= d_{\pm} = \frac{1}{2}(\lambda_1 + \lambda_2)\pm \sqrt{\frac{1}{4}(\lambda_1-\lambda_2)^2+\lambda_5^2} \,, \\
e_1 &= \lambda_3+2\lambda_4-3\lambda_5 \,, \\
e_2 &= \lambda_3 - \lambda_5 \,, \\
f_1 &= f_2 = \lambda_3 + \lambda_4 \,, \\
f_{+} &= \lambda_3 + 2 \lambda_4 + 3 \lambda_5 \,, \\
f_{-} &= \lambda_3  + \lambda_5 \,.
 \end{split}
\end{equation}
where $e_1, e_2, f_+, f_-$ and $f_1$ are the eigenvalues of the submatrix corresponding to the scatterings with initial and final states being one of the states $(\varphi_1^+ \varphi_2^-, \varphi_2^+ \varphi_1^-,h_1 z_2, h_2 z_1, z_1 z_2, h_1 h_2)$. The $a_\pm, b_\pm$ and $c_\pm$ are the eigenvalues of a submatrix corresponding to scatterings with the initial and final states being one of the states: $(\varphi_1^+ \varphi_1^-, \varphi_2^+ \varphi_2^-, z_1 z_1/ \sqrt{2},z_2 z_2/ \sqrt{2},h_1 h_1/ \sqrt{2},h_2 h_2/ \sqrt{2})$.
The solution of the RGEs in Eq. \ref{betafun} is shown in Fig. (\ref{rge}). As shown in this figure the coupling $\lambda_2$ and $\lambda_3$ hit a Landua pole around the scale $4$ TeV. The running of the eigenvalues of the two-particle scattering matrix is depicted in Fig. \ref{unitarity}. For both the couplings $\lambda_i$ and the eigenvalues $a_+, b_+$ and $c_+$, the initial values of the couplings at the top quark mass scale being $173$ GeV, are taken as $\lambda_1\sim 0.51$, $\lambda_2\sim 1.9$, $\lambda_3\sim 4$, $\lambda_4\sim -3$ and $\lambda_5\sim -2.6$. These are a set of couplings that can be taken from Fig. \ref{couprang} so that all the theoretical constraints i.e. the bounded from below condition and the unitarity condition as well as all experimental bounds discussed in section \ref{const} are satisfied.  In fact this set of initial values for the couplings are a benchmark from the viable space in Fig. \ref{2hdm1}. 
It is evident from Fig. \ref{unitarity} that the unitarity is valid up to the scale $4$ TeV. 
All that means that the the current model would be valid up to the scale $4$ TeV. Above this scale another scale invariant model with more degrees of freedom can be envisioned. 

\section{Constraints}\label{const}
{\it Theoretical constraints.}
The bounded from below conditions can be read from the mass eigenvalues being simply
\begin{equation}
\lambda_1>0 \wedge \lambda_2>0 \wedge \lambda_3>-\sqrt{\lambda_1 \lambda_2} \wedge \lambda_5<0 .
\end{equation}
As studied in section \ref{perunit}, generically the perturbativity bounds on the scalar couplings are $\lambda_i \leq 4\pi$ \cite{Ginzburg:2005dt,Chen:2018shg}. Given the results in \cite{Cacchio:2016qyh,Grinstein:2015rtl}, stronger limits are obtained as $\lambda_i \leq 4$ \cite{Atkinson:2021eox}.
Looking at the Yukawa terms involving the charged Higgs couplings to  the quarks and imposing the perturbativity limits as
\begin{equation}
 \frac{\sqrt{2}V_{tb}m_t \cot \beta}{2v}  \leq 4\pi, \hspace{.5cm}
 \frac{\sqrt{2}V_{tb}m_b \tan \beta}{2v}  \leq 4\pi,
\end{equation}
a conservative range is obtained for $\tan \beta$: $ 0.04 \leq \tan \beta \leq 900$ \cite{Atkinson:2021eox}. However, in order to make the comparison between the type I and type II, the upper bound for $\tan \beta$ is taken to be $\sim 300$ \cite{Atkinson:2021eox}.\\

{\it Experimental constraints.}
Constraints on the parameters of the 2HDM by experiments are vastly studied in the literature. 
Here we recap some important limits imposed by various collider data which are applicable in the present work.
The relevant bounds depend on the type of 2HDM under consideration.

The radiative decay ${\bar B} \to X_s \gamma$ is the most important observable to bound the charged Higgs from below in the 2HDM-II. 
In a recent study updating the SM predictions, the strong constraint
on the charged Hiss mass $m_{H^{\pm}} \gtrsim 800$ GeV at $95\%$ CL for 
$\tan \beta \gtrsim 1$ is set\cite{Misiak:2020vlo}. For $\tan \beta \lesssim 1$, the bound becomes even stronger.
The radiative decay alone is not enough to find a concrete lower bound for the charged Higgs mass in 2HDM-I because of the large correction with $\tan \beta$. 
However, based on an estimate taking the contributions to the Wilson coefficients $C_7$ and
$C_8$ at NLO in \cite{Atkinson:2022pcn},
the lower bound on the charged Higgs mass is estimated at $95\%$ CL 
as $m_{H^{\pm}} \gtrsim 300$ GeV for $\tan \beta \gtrsim 1$.

Considering the combination of the tree-level leptonic and semi-leptonic 
decays of $B, B_s, D, D_s, K$ and $\pi$ mesons and the hadronic decays 
of $\tau$ leptons to $\pi$ and $K$ mesons along with $R(D)$ and $R(D^*)$, the viable parameter space is found in the plane $\tan \beta-m_{H^{\pm}}$ for 2HDM-II \cite{Atkinson:2021eox}. The results indicate that for $\tan \beta \lesssim 2$ the lower limit on the charged Higgs mass is $m_{H^{\pm}} \gtrsim 30$ GeV, and the bounds become stronger for larger $\tan \beta$. The same analysis in 2HDM-I finds that $m_{H^{\pm}} \gtrsim 30$ GeV for $\tan \beta \gtrsim 0.5$ at $2\sigma$ \cite{Atkinson:2022pcn}.
In our study we also apply the allowed regions in the $\tan \beta$-$m_{H^{\pm}}$ plane
from the mass difference in the mixing of $B_d$ and $B_s$ mesons in 2HDM-I and 2HDM-II exploiting the results in \cite{Atkinson:2021eox,Atkinson:2022pcn}.

The CMS experiments at the LHC run-II, put strong constraints on the masses of the neutral scalars $h$ and $A$ through $h\to A Z$ and $A\to h Z$ processes in the 2HDM type-II scenario. As reported in \cite{CMS:2016xnc} for $\tan \beta=1.5$ a large portion of the mass parameter space $M_h=200-700$ GeV and $M_A=20-270$ GeV for the decay $h\to A Z$, and $M_A=200-700$ GeV and $M_A=120-270$ GeV for the decay $A\to h Z$ are excluded at the $95 \%$ confidence level (CL). 
An important channel for the CP-odd scalar searches at the ATLAS experiments is the $A\to \tau\tau$ channel, which excludes high $\tan\beta$ regions for $M_A$ as large as $1000$ GeV for 2HDM type-II \cite{ATLAS:2017eiz}.

\section{Results}\label{res}

\begin{figure}
\includegraphics[angle=0,scale=.65]{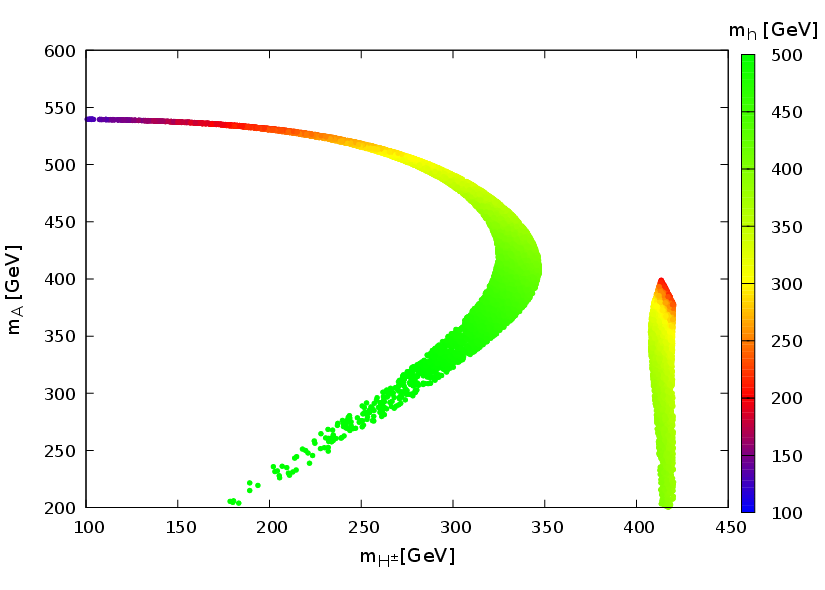}
\caption{The viable range of the new scalar masses is shown when all the theoretical constraints and the bound from global fit of the electroweak data and CDF-II are taken into account.}
\label{massrng}
\end{figure}

\begin{figure}
\includegraphics[angle=0,scale=.65]{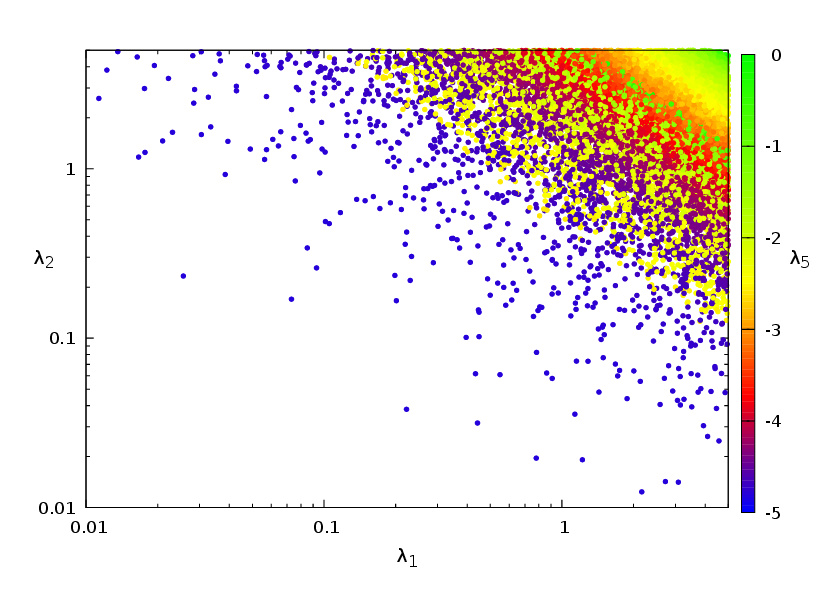}
\caption{The viable range of the quartic couplings is shown when all the theoretical constraints and the bound from global fit of the electroweak data and CDF-II are taken into account.}
\label{couprang}
\end{figure}

The present model has three free parameters, $\lambda_1$, $\lambda_2$ and $\lambda_3$. We scan over the three-dimensional parameter space by sampling the three couplings in the ranges: $0< \lambda_1 < 5$, $0< \lambda_2 < 5$ and $-5< \lambda_3 < 5$, bearing in mind that $\lambda_3 > -\sqrt{\lambda_1 \lambda_2}$ from the vacuum stability conditions.
From Eq. (\ref{diagmass}) two neutral scalar masses $M_h\equiv m_h(v)$ and $M_h\equiv m_A(v)$ are then obtained. Plugging these two masses in Eq. (\ref{mH2}), the mass of the charged Higgs will be determined. For the SM Higgs mass we take  $m_H = 125.25$ GeV. The oblique parameters $S$ and $T$ depend on the three new scalars and we compute them using the relations in Eq. (\ref{op}).   
We will employ the recent global fit of electroweak data incorporating the CDF-II measurements of the $W$-boson mass, estimated in \cite{Strumia:2022qkt} as $S_0$ and $T_0$ while fixing $U \sim 0$. 
First, we are interested in finding the viable values for the new masses respecting all the theoretical constraints as well as the relation in Eq. (\ref{STconst})
at $95\%$ CL equivalent to $R^2=5.99$. Our numerical results in Fig. \ref{massrng} shows the range of the masses for the 
scalars, $H^{\pm}$, $M_h$ and $M_A$. The largest mass allowed for the charged Higgs 
is about $430$ GeV, while the mass of other two scalars can reach about $500$ GeV. Fig. \ref{couprang}  shows the viable regions for the quartic couplings $\lambda_1, \lambda_2$ and $\lambda_5$ which as in Fig. \ref{massrng}, satisfy
the bounded from below and the unitarity conditions, give the correct value for the scalon mass to be $125$ GeV, and on top of that explain the $W$-boson mass anomaly. Note that the coupling $\lambda_3$ is fixed from Eq. (\ref{mH2}).

\begin{figure}
\includegraphics[scale=.65]{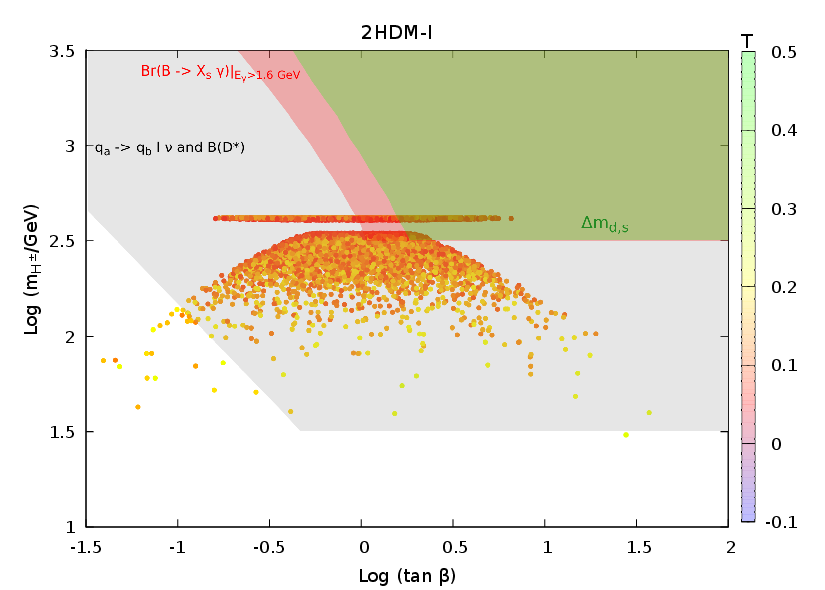}
\caption{Shown is the allowed regions from various collider experiments colored in Red, Green and Gray in $\tan \beta-m_{H^{\pm}}$ plane. There is a viable space of parameters for the SI-2HDM type I where the $W$-boson mass anomaly is explained and all the collider constraints are satisfied.}
\label{2hdm1}
\end{figure}

\begin{figure}
\includegraphics[scale=.65]{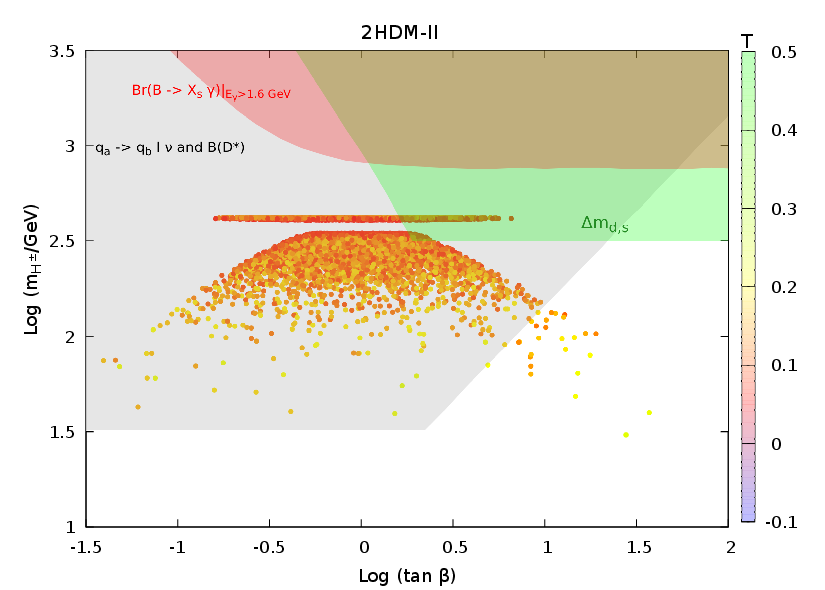}
\caption{Shown is the allowed regions from various collider experiments colored in Red, Geen and Gray in $\tan \beta-m_{H^{\pm}}$ plane. The parameter space that explains the $W$-boson mass anomaly, does not overlap the regions allowed by all collider constraints, therefore the SI-2HDM type II is excluded.}
\label{2hdm2}
\end{figure}

Next we impose experimental bounds recounted in the previous section, on the restricted ranges of the charged Higgs mass. We consider both 2HDM-I and 2HDM-II in our analysis.
The results concerning 2HDM-I and 2HDM-II in the $\tan \beta$-$m_{H^{\pm}}$ plane  
are shown in Fig. \ref{2hdm1} and Fig. \ref{2hdm2}, respectively. The three different experimental bounds do not overlap when 2HDM type-II is 
considered, indicating that 2HDM-II in our scenario is excluded by the experiments at $2\sigma$. 
The experimental bounds are weaker in the case we choose 2HDM-I.  In fact, we see from Fig. \ref{2hdm1} that a large part of a patch in the $\tan \beta$-$m_{H^{\pm}}$ plane with charged Higgs mass in the range 
$385$ GeV $\lesssim m_{H^{\pm}} \lesssim 440$ GeV and $1.6 \lesssim \tan \beta \lesssim 6.8$ remains viable. In addition,  there is a small viable region 
around $m_{H^{\pm}}\sim 320-360$ GeV with $\tan \beta \sim 1.8-2.5$.  

The CMS constraints from the decays $A\to hZ$ and $h\to AZ$ on the neutral scalars masses are weak in the 2HDM-I scenario, therefore they do not change the viable space found in Fig. \ref{2hdm1}. As seen in Fig. \ref{2hdm2}, also the ATLAS bound on the decay $A\to \tau^+ \tau^-$, does not constrain the parameter space of the 2HDM-II. The exclusion from this channel for the type I is even weaker and therefore irrelevant here. 

\section{Conclusion}\label{conc}
The recent CDF-II collaboration report on the measurement on the $W$ gauge boson mass could be a window to New Physics (NP) if it gets confirmed by another experiment e.g. at the LHC. The $M_W$ reported by CDF-II is significantly larger than what the Standard Model (SM) predicts. Therefore, to explain the anomaly one needs extra degrees of freedom that affects the mass of the $W$ boson. One choice is to consider the SI-2HDM in which the Higgs particle is the classically massless scalon which becomes massive only by radiative corrections {\it \`{a} la} Coleman and Weinberg. The scale symmetry makes the model quite restrictive so that it becomes more predictive respect to the generic 2HDM. Therefore, in the SI-2HDM we deal with less number of free parameters. Apart from the SM Higgs, there are two charged scalar components $H^\pm$, a CP-even scalar $h$, and a CP-odd scalar $A$. The oblique parameters defined by Peskin and Takeuchi, depend on the masses $M_{H^\pm}$, $M_h$ and $M_A$. We have shown that in the SI-2HDM, the desired $S$ and $T$ oblique parameters can become large enough to explain the excess in the mass of the $W$ boson. The allowed masses of the charged and neutral scalars to accommodate the $W$ mass anomaly are shown in Fig. \ref{massrng}. We have also imposed constraints on the charged Higgs mass and the neutral scalars from colliders. In Fig. \ref{2hdm1}, the bounds are provided for the 2HDM-I and in Fig. \ref{2hdm2} for the 2HDM-II. The outcome of imposing these experimental constraints on the SI-2HDM is that the 2HDM type-II is completely excluded by the charged Higgs mass bounds, while for the 2HDM-I there are viable regions. 

\section*{Acknowledgment}
P.G. is grateful to the HECAP division at ICTP for the hospitality and support during his visit when this work was done.
\bibliographystyle{apsrev4-1}
\bibliography{ref.bib}
\end{document}